\documentclass[sigconf,screen]{acmart}

\usepackage{tikz}
\usetikzlibrary{tikzmark}
\usetikzlibrary{patterns}
\usepackage{xspace}
\usepackage{multirow}
\usepackage{balance}
\usepackage{ctable}
\usepackage{listings}
\usepackage{color}
\usepackage{subcaption}
\usepackage{graphicx}
\graphicspath{ {figures/} }
\usepackage{url}
\usepackage[justification=centering]{caption}
\usepackage{algorithm}
\usepackage{algorithmicx}
\usepackage{algpseudocode}
\usepackage{amssymb}
\usepackage{amsmath}
\usepackage{amsfonts}
\usepackage{newfloat}
\usepackage{cleveref}
\usepackage{acronym}
\usepackage{verbatim} 
\usepackage{xcolor}
\usepackage[nounderscore]{syntax}
\usepackage{multicol}
\usepackage{multirow}
\usepackage[flushleft]{threeparttable}
\usepackage{enumitem}
\usepackage{array, makecell, caption}
\usepackage{cellspace}

\newcommand{\ie}{i.e.,\xspace}
\newcommand{\eg}{e.g.,\xspace}

\newcommand{\RQ}[2]{\vspace{3pt}\noindent\textbf{#1:} #2}

\newfloat{program}{h}{eqn}
\newlength{\MaxSizeOfLineNumbers}%
\definecolor{keywordcolor}{rgb}{0.8,0.1,0.5}
\definecolor{lightlightgray}{gray}{.96}
\definecolor{lightgray}{gray}{.925}
\definecolor{medlightgray}{gray}{0.7}
\definecolor{medgray}{gray}{0.4}
\definecolor{darkgray}{gray}{0.35}
\definecolor{nearblack}{gray}{0.15}

\newcommand{\code}[1]{{\small\texttt{#1}}}

\crefname{program}{Program}{Programs}

\definecolor{javared}{rgb}{0.6,0,0} 
\definecolor{javagreen}{rgb}{0.25,0.5,0.35} 
\definecolor{javapurple}{rgb}{0.5,0,0.35} 
\definecolor{javadocblue}{rgb}{0.25,0.35,0.75} 
\definecolor{pastelyellow}{rgb}{0.99, 0.99, 0.59}
\definecolor{peach-orange}{rgb}{1.0, 0.8, 0.6}
\definecolor{peach}{rgb}{1.0, 0.9, 0.71}
\definecolor{lightgoldenrodyellow}{rgb}{0.98, 0.98, 0.82}
\definecolor{lightgreen}{HTML}{CCFFCC}
\definecolor{magicmint}{rgb}{0.67, 0.94, 0.82}
\definecolor{lightmauve}{rgb}{0.86, 0.82, 1.0}
\definecolor{palepink}{rgb}{0.98, 0.85, 0.87}
\definecolor{lightapricot}{rgb}{0.99, 0.84, 0.69}
\definecolor{blizzardblue}{rgb}{0.67, 0.9, 0.93}
\definecolor{mauve}{HTML}{E0D7FF}
\definecolor{orange}{HTML}{FFE5B4}
\definecolor{blue}{HTML}{BFEFFF}
 
\acrodef{ast}[AST]{Abstract Syntax Tree}
\acrodef{abstract-tree}[APG]{Abstract Program Graph}
\acrodef{localization}[FSL]{Failure Scenario Localization}
\acrodef{pattern}[ESP]{Exception Scenario Pattern}

\newcommand{\toolname}{\mbox{\sc Maestro}\xspace}
\newcommand{\so}{\mbox{Stack Overflow}\xspace}

\newcommand{\prompter}{\mbox{\sc Prompter}\xspace}

\acrodef{so}[SO]{\mbox{Stack Overflow}}
\acrodef{re}[RE]{runtime exception}

\newcommand{\soQuesNum}{18 million\xspace}
\newcommand{\soAnsNum}{28 million\xspace}
\newcommand{\soVisitNum}{260 million\xspace}
\newcommand{\numSoJavaPosts}{2.65 million\xspace}
\newcommand{\numSoREPosts}{193,186\xspace}
\newcommand{\numJavaRETypes}{53\xspace}
\newcommand{\benchmarkSize}{78\xspace}
\newcommand{\maestroIHScore}{71\%\xspace}
\newcommand{\otherIHScoreLow}{8\%\xspace}
\newcommand{\otherIHScoreHigh}{44\%\xspace}
\newcommand{\uStudyMaestroIHScore}{80\%\xspace}

\AtBeginDocument{%
  \providecommand\BibTeX{{%
    \normalfont B\kern-0.5em{\scshape i\kern-0.25em b}\kern-0.8em\TeX}}}

\setcopyright{acmcopyright}
\acmPrice{15.00}
\acmDOI{10.1145/3368089.3409764}
\acmYear{2020}
\copyrightyear{2020}
\acmSubmissionID{fse20main-p894-p}
\acmISBN{978-1-4503-7043-1/20/11}
\acmConference[ESEC/FSE '20]{Proceedings of the 28th ACM Joint European Software Engineering Conference and Symposium on the Foundations of Software Engineering}{November 8--13, 2020}{Virtual Event, USA}
\acmBooktitle{Proceedings of the 28th ACM Joint European Software Engineering Conference and Symposium on the Foundations of Software Engineering (ESEC/FSE '20), November 8--13, 2020, Virtual Event, USA}

\begin{document}

\title{Recommending Stack Overflow Posts for Fixing Runtime Exceptions using Failure Scenario Matching}

\author{Sonal Mahajan}
\affiliation{
\institution{Fujitsu Laboratories of America, Inc.}
\city{Sunnyvale}
\country{USA}}
\email{smahajan@fujitsu.com}

\author{Negarsadat Abolhassani}
\authornote{This work was done when the author was an intern at Fujitsu Laboratories of America}
\affiliation{
\institution{University of Southern California}
\city{Los Angeles}
\country{USA}}
\email{abolhass@usc.edu}

\author{Mukul R. Prasad}
\affiliation{
\institution{Fujitsu Laboratories of America, Inc.}
\city{Sunnyvale}
\country{USA}}
\email{mukul@fujitsu.com}

\renewcommand{\shortauthors}{Mahajan, et al.}

\begin{abstract}
Using online Q\&A forums, such as \ac{so}, for guidance to resolve program bugs, among other development issues, is commonplace in modern software development practice. Runtime exceptions (RE) is one such important class of bugs that is actively discussed on \ac{so}. In this work we present a technique and prototype tool called \toolname that can automatically recommend an \ac{so} post that is most relevant to a given Java \acs{re} in a developer's code. \toolname compares the exception-generating program scenario in the developer's code with that discussed in an \ac{so} post and returns the post with the closest match. To extract and compare the exception scenario effectively, \toolname first uses the answer code snippets in a post to implicate a subset of lines in the post's question code snippet as responsible for the exception and then compares these lines with the developer's code in terms of their respective \ac{abstract-tree} representations. The \acs{abstract-tree} is a simplified and abstracted derivative of an abstract syntax tree, proposed in this work, that allows an effective comparison of the functionality embodied in the high-level program structure, while discarding many of the low-level syntactic or semantic differences. We evaluate \toolname on a benchmark of \benchmarkSize instances of Java \acs{re}s extracted from the top 500 Java projects on GitHub and show that \toolname can return either a highly relevant or somewhat relevant \ac{so} post corresponding to the exception instance in \maestroIHScore of the cases, compared to relevant posts returned in only \otherIHScoreLow \,- \otherIHScoreHigh instances, by four competitor tools based on state-of-the-art techniques.  We also conduct a user experience study of \toolname with 10 Java developers,
where the participants judge \toolname reporting a highly relevant or somewhat relevant post in \uStudyMaestroIHScore of the instances. In some cases the post is judged to be even better than the one manually found by the participant.
\end{abstract}



\maketitle

\section{Introduction}
\label{sec:introduction}
Software developers regularly refer to online Q\&A forums for a wide variety of development tasks, from system design and configuration, to code completion, to software debugging and patching~\cite{Brandt:2009, Sadowski:2015, DBLP:journals/corr/abs-1802-02938, Wu2019}. \acf{so}, the most popular such forum, with over \soAnsNum answers to \soQuesNum questions, sees nearly \soVisitNum views per month~\cite{SOInfo}. Software debugging and patching is a resource-intensive development activity, consuming up to $50\%$ of developers' time~\cite{bug-cost-study:2013}. In particular, \acfp{re} are an important class of bugs that have been recognized as having a severe impact on system availability and crashes~\cite{Li:ASID06}. Realizing their importance, researchers have proposed automated debugging, repair, and recovery techniques specifically addressing runtime errors in general and \acp{re} in particular~\cite{Sinha:ISSTA2009, NPEFix:arXiv2015, QACrashFix:ASE2015, Genesis:FSE2017, VFix:ICSE2019, Long:PLDI2014, Ares:ASE2016}. 
In fact, of the nearly \numSoJavaPosts posts on \ac{so} tagged with \code{Java} and/or \code{Android} nearly \numSoREPosts~\footnote{Obtained by searching \ac{so} with the name of any of the top \numJavaRETypes  most common Java \acp{re}.} -- a remarkable 7\% -- are related just to Java \acp{re}, 
showing that developers are actively discussing Java \acp{re} on \ac{so}, presumably to resolve such errors in their own code.

In this paper we present a technique and prototype tool \toolname (\textbf{M}ine and \textbf{A}nalyz\textbf{E} \textbf{ST}ackoverflow to fix \textbf{R}untime excepti\textbf{O}ns) that can automatically find an \ac{so} post that is \emph{most relevant} to a given \ac{re} in a developer's (Java) code. We define the most relevant post (or posts, since there could be several) as one discussing the same runtime scenario, exciting the same type of \ac{re} as the developer's code. Such a tool can save the time it would take the developer to understand the high-level scenario producing the exception, create a search query based on it, search \ac{so} with the query and manually browse the search results, one discussion post at a time, to identify the most relevant post. These steps may need to be repeated several times till an acceptable post is found. 
Once found, the user could potentially use one of the suggested answers to fix their bug. This fixing could be done manually or assisted by a tool like {\sc ExampleStack}~\cite{ExampleStack:ICSE2019}.

One solution to the above problem is to use \ac{so} post recommendation techniques like \prompter~\cite{Prompter:MSR2014, Libra:ICSE2017, Seahawk:CSMR2013}, which search for a post most relevant to the user's code context, \ie the \emph{function} the user's code is trying to implement. However, as shown in Section~\ref{sec:evaluation}, these techniques do not work for our problem. The reason is that what we seek is \emph{not} a post discussing the overall function the user's code was implementing but rather one addressing the specific sequence of program state manipulations that raised the exception. The overall function of the method within which these manipulations occur is somewhat irrelevant. Another idea would be to use traditional code clone detection techniques, such as \cite{DECKARD:ICSE2007, SourcererCC:ICSE2016, NiCad:ICPC2011, CCFinder:TSE2002}, or code-to-code search techniques, such as \cite{FaCoY:ICSE2018, AROMA:OOPSLA2019} to check correspondence between the user's code and the question code snippet in a given \ac{so} post. However, this approach would not work either (see evaluation in Section~\ref{sec:evaluation}), in part because of the above reason -- it is not apparent what portions to try to match between the exception-throwing developer code and the \ac{so} code snippet. Further, even the relevant lines of the \ac{so} post snippet, if identified, could differ significantly from their counterparts in the developer code in terms of not only variable names (which code clone detectors easily handle) but also data types and program constructs (\eg a \code{while} vs. a \code{for} loop), while instantiating the same core exception-causing scenario. Thus, this problem seems outside the realm of traditional code clone detection or code search, since the matching criterion is the cause of the exception, rather than the function of the containing method.

\begin{figure*}[t]
\centering
\begin{subfigure}[b]{0.9\textwidth}
\begin{lstlisting}[language=Java, firstnumber=211, mathescape, framexleftmargin=2em, basicstyle=\fontsize{7pt}{7pt}\selectfont\ttfamily, numberstyle=\scriptsize]
public void dropColumn(String databaseName, String tableName, String columnName)
{
    verifyCanDropColumn(this, databaseName, tableName, columnName);
    org.apache.hadoop.hive.metastore.api.Table table = delegate.getTable(databaseName, tableName)
            .orElseThrow(() -> new TableNotFoundException(new SchemaTableName(databaseName, tableName)));
		 @\colorbox{mauve}{\strut \textbf{for} (FieldSchema fieldSchema : table.getSd().getCols())}@ { @\hspace{5pt} {\normalsize \textbf{\color{red} $\longleftarrow$ exception thrown here}}@
        @\colorbox{orange}{\strut \textbf{if} (fieldSchema.getName().equals(columnName))}@ {
            @\colorbox{lightgreen}{\strut table.getSd().getCols().remove(fieldSchema);}@
        }
    }
    alterTable(databaseName, tableName, table);
}
\end{lstlisting}
\vspace{-5pt}
\caption{\label{fig:buggy-code}Buggy source code (BridgingHiveMetastore.java)}
\end{subfigure}
\par\bigskip
\begin{subfigure}[b]{0.9\columnwidth}
\begin{lstlisting}[language=Java, basicstyle=\fontsize{7pt}{7pt}\selectfont\ttfamily, numberstyle=\scriptsize]
public static void main(String[] args) {
    User user = new User("user1","user1",1l);
    User user1 = new User("user2","user2",2l);
    User user2 = new User("user3","user3",3l);

    List<User> list = new ArrayList<User>();
    list.add(user);
    list.add(user1);
    list.add(user2);

    @\colorbox{mauve}{\strut \textbf{for} (User user3 : list)}@ {
        System.out.println(user3.getName());
        @\colorbox{orange}{\strut \textbf{if} (user3.getName().equals("user1"))}@ {
            @\colorbox{lightgreen}{\strut list.remove(user3);}@
        }
    }
} @\hspace*{\fill}{\large \textbf{Q}}@
\end{lstlisting}
\vspace{-5pt}
\caption{\label{fig:so-post1-ques}Stack Overflow post \#21973342 question code snippet}
\end{subfigure}
\quad
\begin{subfigure}[b]{1.1\columnwidth}
\begin{subfigure}[b]{\columnwidth}
\begin{lstlisting}[language=Java, basicstyle=\fontsize{7pt}{7pt}\selectfont\ttfamily, numberstyle=\scriptsize]
Iterator<User> it = list.iterator();
@\colorbox{mauve}{\strut \textbf{while} (it.hasNext())} @{
  User user = it.next();
  @\colorbox{orange}{\strut \textbf{if} (user.getName().equals("user1"))} @{
      @\colorbox{lightgreen}{\strut it.remove();}@
  }
} @\hspace*{\fill}{\large \textbf{A}}@
\end{lstlisting}
\vspace{-5pt}
\caption{\label{fig:so-post1-ans}Stack Overflow post \#21973342 answer code snippet}
\end{subfigure}
\par\bigskip
\begin{subfigure}[b]{\columnwidth}
\begin{lstlisting}[language=Java, basicstyle=\fontsize{7pt}{7pt}\selectfont\ttfamily, numberstyle=\scriptsize]
Collection<T> myCollection; ///assume it is initialized and filled
for (Iterator<?> index = myCollection.iterator(); index.hasNext();) {
    Object item = index.next();
    myCollection.remove(item);
} @\hspace*{\fill}{\large \textbf{Q}}@
\end{lstlisting}
\vspace{-5pt}
\caption{\label{fig:so-post2-ques}Stack Overflow post \#2054189 question code snippet}
\end{subfigure}
\end{subfigure}
\caption{\label{fig:example}Example from Presto (\url{github.com/prestodb/presto}) throwing ConcurrentModificationException (issue \#9733)}
\end{figure*}

Our approach is designed around three key insights. First, in most \ac{so} discussion posts the question includes a code snippet exemplifying the scenario being discussed. In particular, for posts addressing \acp{re}, the question code snippet naturally includes a \emph{structured} description of the exception-raising scenario. Thus, our approach exclusively uses this question code snippet and compares it to the exception-throwing user code to decide the relevancy of the post. Second, the question code snippet may include code to make the snippet functionally or syntactically complete, but otherwise irrelevant to the exception scenario. However, the answers in the post also include code snippets, suggesting solutions to the discussed problems. In our case these answer code snippets often suggest patches for the \ac{re} and often more directly address the failure producing lines. Thus, we use the answer code snippets to identify the lines in the question code snippet relevant to the exception scenario and discard the rest. We term this \textit{\ac{localization}}. Third, as mentioned above, the failure producing lines in the developer code could have substantial syntactic differences from those the \ac{so} code snippet. To facilitate a meaningful comparison we develop a representation called an \textit{\acf{abstract-tree}}. Conceptually, an \ac{abstract-tree} is a simplified, abstracted, and (partially) canonicalized derivative of the \ac{ast} for a piece of code. Thus, we compare the developer code and question code snippet (after performing \ac{localization}) in terms of their respective \acp{abstract-tree}. We also use the \ac{abstract-tree} representation when aligning answer code snippets in a post with the question code snippet for the purpose of \acl{localization}.

We develop a technique and prototype tool, \toolname based on the above insights and conduct an internal evaluation on a benchmark of \benchmarkSize instances of Java \acp{re}, spanning 19 exception types, extracted from the repositories of the top 500 Java projects on GitHub. The evaluation shows that \toolname returns either a highly relevant or somewhat relevant \ac{so} post corresponding to the exception instance for \maestroIHScore of the instances, compared to relevant posts returned in only \otherIHScoreLow \,- \otherIHScoreHigh instances, by four competitor tools based on state-of-the-art techniques. Further, a comparison with three different baseline variations of \toolname shows that each of the key design features of \toolname are essential to its overall performance. We also conduct a user experience study of \toolname with 10 Java developers, where the participants judge \toolname as reporting a highly relevant or somewhat relevant post in \uStudyMaestroIHScore of the instances. In some cases the post is judged to be even better than the one manually found by the participant.

This main contributions of this paper are as follows:
\begin{itemize}[leftmargin=10pt]
    \item \textbf{Technique:} An automated technique, \toolname for recommending relevant \so posts to Java developers which could assist them in diagnosing and fixing \acp{re} in their code.
    \item \textbf{Tool:} A prototype implementation of the \toolname technique along with four different baseline variations of it, to evaluate its design features.
    \item \textbf{Evaluation:} An evaluation of \toolname, its three baseline variants, and four competitor tools, on a benchmark of \benchmarkSize \ac{re} instances extracted from the top 500 Java projects on GitHub.  
    \item \textbf{User experience study:} A user study with 10 Java developers to qualitatively evaluate the performance of \toolname.
\end{itemize}

\section{Illustrative Example}
\label{sec:motivating-example}

\begin{figure*}[t]
\centering
\begin{subfigure}[t]{0.49\textwidth}
\includegraphics[width=\textwidth]{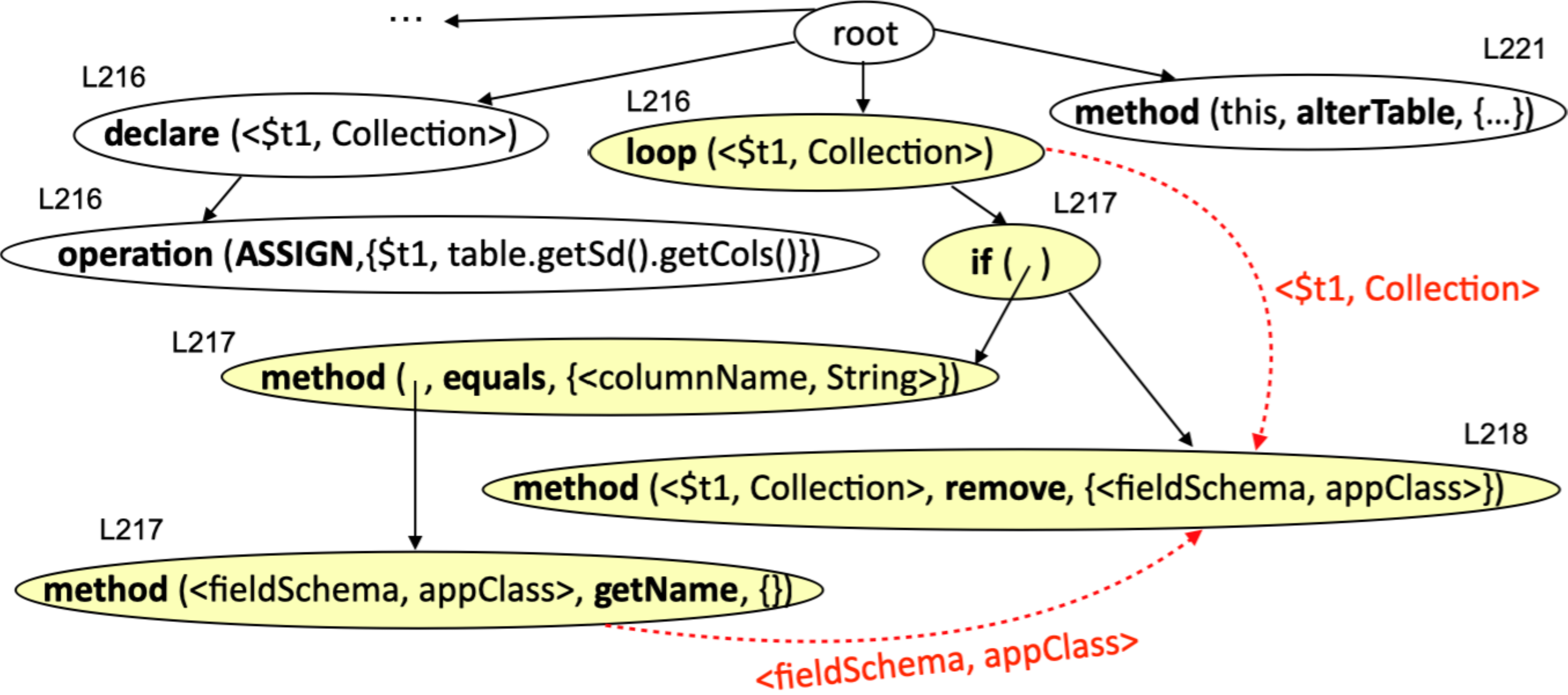}
\caption{\label{fig:buggy-code-abstract-tree} \textit{\ac{abstract-tree}}$_{B}$: \ac{abstract-tree} of buggy code shown in \Cref{fig:buggy-code}}
\end{subfigure}
\begin{subfigure}[t]{0.49\textwidth}
\includegraphics[width=\textwidth]{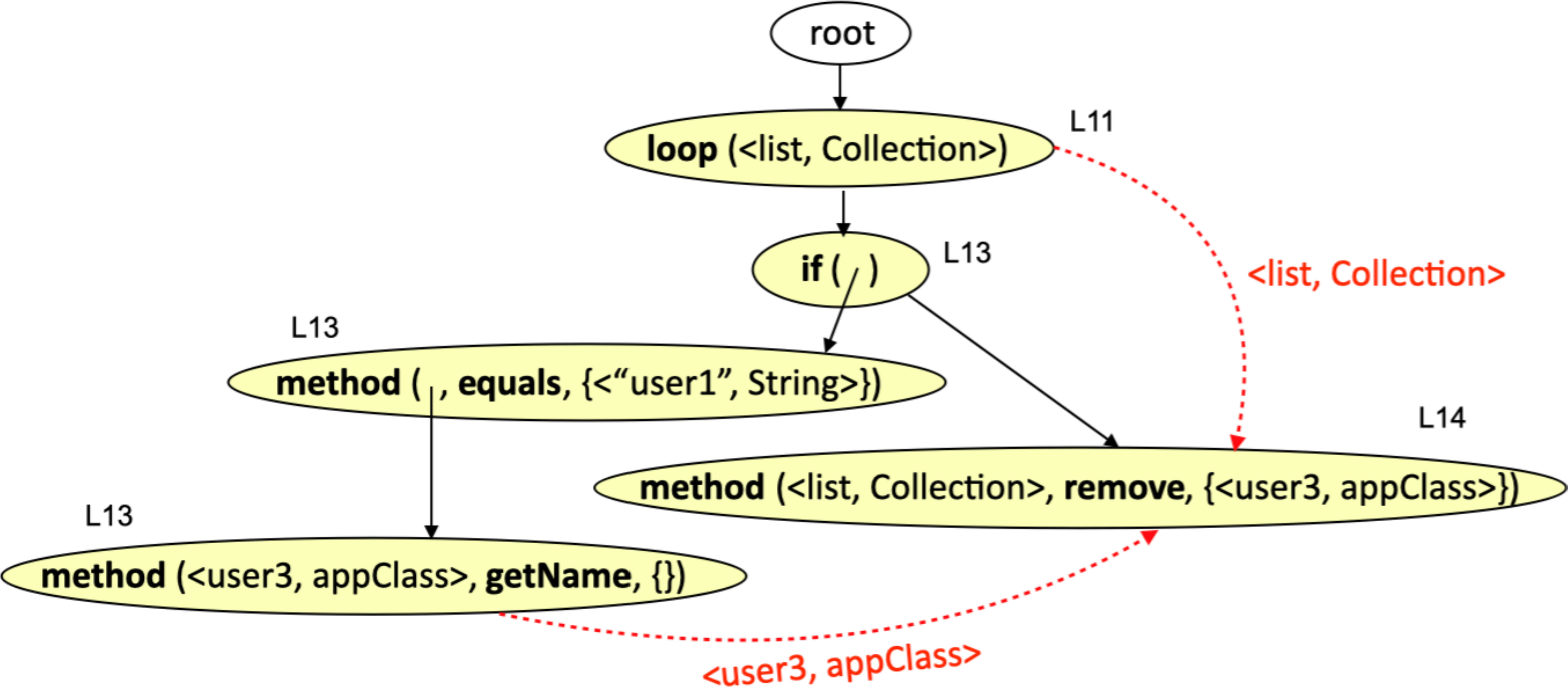}
\caption{\label{fig:so-code-abstract-tree} \textit{\ac{abstract-tree}}$_{Q}$: \ac{abstract-tree} of \so post shown in \Cref{fig:so-post1-ques}}
\end{subfigure}
\caption{\label{fig:graphs} \acp{abstract-tree} of buggy code and \so post }
\end{figure*}

In this section, we illustrate our technique using the example shown in \Cref{fig:example}. Consider the developer's buggy source code given in \Cref{fig:buggy-code}, extracted from the \code{BridgingHiveMetastore.java} file of the Presto project~\cite{presto}, a popular distributed SQL query engine. The buggy source code throws a "\code{ConcurrentModifcationException}", a common \acf{re} that may be thrown by methods detecting concurrent modification on an object, when such modification is not permitted. In this example, the exception is thrown because the \code{Collection} object returned by "\code{table.getSd().getCols()}" is modified at line 218 while being iterated over at line 216.

In order to resolve the encountered \ac{re}, the developer may refer to \acf{so} to find a post discussing the exception in the same context. To facilitate this, \toolname collects \ac{so} posts discussing \code{ConcurrentModificationException}
and compares code snippets in the question section of each post with the developer's buggy code. Figures~\ref{fig:so-post1-ques} and \ref{fig:so-post2-ques} show question code snippets extracted from two \ac{so} posts discussing this \ac{re}.

Let us consider the snippet shown in \Cref{fig:so-post1-ques}. Comparing the snippets from \Cref{fig:buggy-code} and \Cref{fig:so-post1-ques} directly is challenging since their function is different. \Cref{fig:buggy-code}'s function is to drop a database column, while \Cref{fig:so-post1-ques}'s function is to delete a user. The code snippets also have lines that are irrelevant to the exception scenario. In particular, only lines 216-218 from \Cref{fig:buggy-code} and lines 11, 13 and 14 from \Cref{fig:so-post1-ques} are pertinent to the \ac{re} being thrown. To address this challenge, our approach is to extract the failure scenario on the \ac{so} side and then check if it is also instantiated in the buggy code. 

\toolname automatically localizes the failure scenario in the question snippet (\Cref{fig:so-post1-ques}) by leveraging the answer code snippets, such as the one shown in \Cref{fig:so-post1-ans}. The answer snippet in \Cref{fig:so-post1-ans} suggests a way for fixing the \code{ConcurrentModificationException} by using "\code{Iterator}" to remove the user object. The answer snippet clearly points to lines constituting the exception scenario in the question snippet. However, it is challenging to establish this correspondence since the question and answer snippets may include syntactic differences. In this case, the question iterates using a \code{for} loop over the \code{List} object, while the answer iterates using a \code{while} loop over an \code{Iterator} object. To address this challenge, \toolname abstracts and canonicalizes the snippets in the \acf{abstract-tree}, and then structurally aligns the two \acp{abstract-tree} to find the corresponding lines automatically. The color-coding in \Cref{fig:so-post1-ques} and \Cref{fig:so-post1-ans} show the corresponding lines obtained from the \ac{abstract-tree} alignment. \Cref{fig:so-code-abstract-tree} shows the \ac{abstract-tree} representing the matched lines extracted from the question snippet. This \ac{abstract-tree} forms the \acf{pattern} for \code{ConcurrentModificationException} as per this post. Let us call this pattern \textit{\ac{abstract-tree}}$_{Q}$. 

\begin{figure*}[t]
  \centering
  \includegraphics[width=\textwidth]{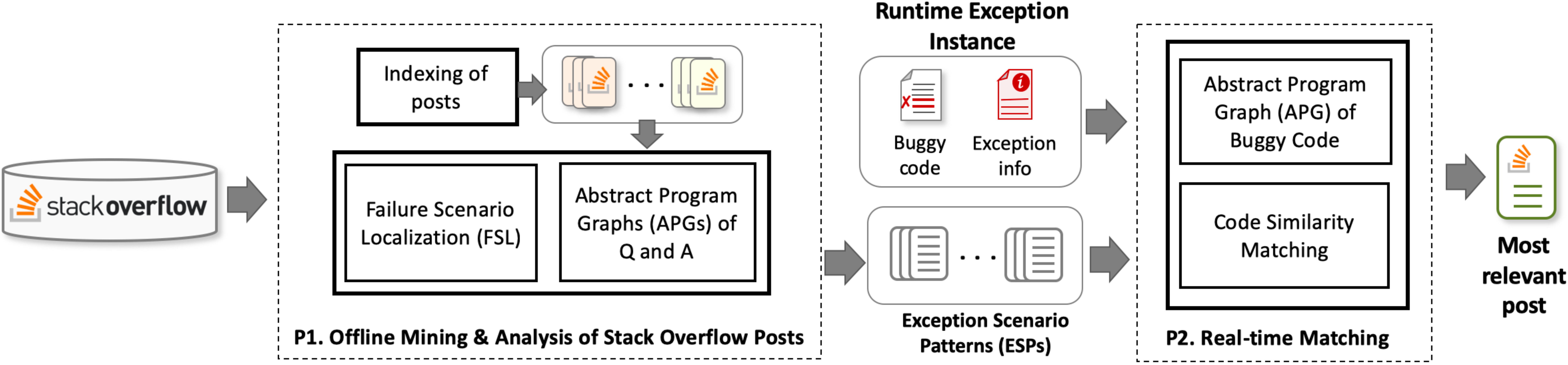}
  \caption{Overview of the approach\label{fig:overview}}
\end{figure*}

To find if this post matches the exception scenario in the developer's code, \toolname first represents the buggy code as an \ac{abstract-tree} (\Cref{fig:buggy-code-abstract-tree}). Let us call it \ac{abstract-tree}$_B$. 
Now the next step is to compare \textit{\ac{abstract-tree}}$_{Q}$ to \textit{\ac{abstract-tree}}$_{B}$. For this, \toolname aligns the two \acp{abstract-tree}, by computing a 1-to-1 node mapping, and computes a similarity score. 
The similarity score is calculated as a weighted sum of the structural and data equivalence. Structural equivalence is comprised of analyzing the programming "\textit{constructs}" used in a matched pair, and if the constructs match then the equivalence between the involved data "\textit{types}" is checked. For both construct equivalence and type equivalence, \toolname applies some level of abstraction depending on program constructs. For example, \toolname identifies user-defined classes as "appClass". Hence, The methods \textbf{remove} in lines 218 and 14 of \Cref{fig:buggy-code} and \Cref{fig:so-post1-ques}, respectively, are equivalent at both type and construct level since both can be abstracted as \code{Collection.\textbf{remove}(appClass)}. The results for \ac{abstract-tree} alignment is shown in \Cref{fig:buggy-code-abstract-tree} and \Cref{fig:so-code-abstract-tree}. The nodes exhibiting both construct and type similarity are shown in yellow. The red arrows in the \acp{abstract-tree} of \Cref{fig:graphs} show the data edges. Data similarity is calculated as the similarity between the uses of variables appearing in the equivalent nodes of \textit{\ac{abstract-tree}}$_{Q}$ and \textit{\ac{abstract-tree}}$_{B}$. For example, both variables $\$t1$ and $list$ in the matched \code{\textbf{loop}} nodes from \Cref{fig:buggy-code-abstract-tree} and \Cref{fig:so-code-abstract-tree}, respectively, are used in the matched \code{\textbf{remove}} nodes as well. Therefore, this example has five pairs of construct \textit{and} type-matched nodes, and a perfect match between the uses of the two variables $\$t1$ and $list$. Assuming the weights for construct, type, and data similarity to be 1, 2, and 0.5, respectively, the similarity score for this post (\Cref{fig:so-post1-ques}) is 16.

\Cref{fig:so-post2-ques} shows a code snippet from another \so post. Assuming the same weights as above, the similarity score for this is 2.5, since there only two construct matches for \textbf{\code{loop}} and \textbf{\code{remove}} at lines 2 and 5, respectively, no type matches, and only one var-use match ($\$t1$). Therefore, \toolname chooses the first post (\Cref{fig:so-post1-ques}) with score 16 as the most relevant post for the developer's buggy code, over the second post (\Cref{fig:so-post2-ques}) with score 2.5. This result is correct since the first post very closely mimics the exception scenario of the developer than the second post.

\section{Approach}
\label{sec:approach}

The goal of our approach is to automatically find the most relevant \acf{so} post for the \acf{re} encountered by the developer. Most \ac{so} posts discussing \ac{re}s include program artifacts, such as a code snippet exemplifying the exception scenario being discussed, or a failure stack trace. These artifacts could potentially be used as a basis for establishing relevance to the developer's failure. However, stack traces appear in a relatively small fraction of posts. Therefore, we decided to design our approach around matching of code snippets. Specifically, our key insight in finding the most relevant post is that if the developer's code exhibits the \textit{same exception scenario} as presented in the code snippet in the \ac{so} question, then it is very likely that the post will contain answers to fix the exception in the developer's code context.

Finding a relevant post based on exception scenario similarity is complicated by several challenges. The first challenge is in capturing the exception scenario in the \ac{so} post. Code snippets in the question section of \ac{so} posts can be arbitrarily large, while the statements capturing the \textit{exception scenario}, \ie the lines that are responsible for exciting the \ac{re} are typically small. For example, the question code in \Cref{fig:so-post1-ques} is of 17 lines, while the lines capturing the scenario of \code{ConcurrentModificationException} are only three (lines 11, 12, and 13). This challenge requires that our solution correctly identify the lines relevant to the \ac{re} under consideration. The second challenge is that the developer's code and the \ac{so} code may have significant syntactic and semantic differences, such as in the structure of the code (\eg interleaving of the exception scenario with added/removed lines), syntactic constructs (\eg \texttt{for} vs. \code{while}), and identifier names. This challenge requires that our solution tolerates the various syntactic and semantic differences between the developer's code and \ac{so} code while establishing similarity.

Two insights into these challenges guide the design of our approach. The first insight is that it is possible to automatically identify the potentially relevant lines contributing to the exception scenario in \ac{so} question code snippets by referring to the answers of that question. Code snippets in \ac{so} answers tend to be more pointed in discussing the problem in the question code while suggesting ways of fixing it. Leveraging this observation, we designed a \textit{\acf{localization}} technique that overlays answer code  on question code to identify the relevant lines (\Cref{sec:localization}). The second insight is that it is possible to abstract out the syntactic and semantic differences allowing for a normalized structural code comparison between developer's code and \ac{so}  code. For this, we designed the \textit{\acf{abstract-tree}} that captures the structural and data relationships between program statements, while abstracting out low-level syntactic details (\Cref{sec:abstract-tree}). Further, our \textit{matching} function calculates a weighted similarity score (\Cref{sec:matching}).

Our approach can be roughly broken down into two distinct phases, \textit{offline mining and analysis} of \ac{so} posts to index them by \acfp{pattern} and \textit{real-time matching} of the mined \ac{so} posts and developer's buggy code to find the most relevant post. These are shown in \Cref{fig:overview}. The first phase selects and then indexes \ac{so} posts related to \acp{re}.
For each indexed post, it then performs \ac{localization} to identify the lines of code relevant to the \ac{re} type being discussed in the post, by using answer pointers. This involves representing the \ac{so} post's question and answer code snippets as \acp{abstract-tree} and then structurally aligning (\Cref{sec:alignment}) them to find the overlapping nodes. The \ac{abstract-tree} of the question is then pruned by removing the nodes not matched to the answer \ac{abstract-tree}, and saved in persistent storage for use in the second phase. We refer to the pruned \ac{abstract-tree} as the \textit{\ac{pattern}}. The second phase takes three inputs: \ac{re} information (name and failing line available from the stack trace), buggy source code containing the failing line, and the \acp{pattern} extracted from the first phase. The second phase begins by representing the buggy code as an \ac{abstract-tree}. It then performs a \textit{code similarity matching} between the \ac{abstract-tree} of the buggy code and each of the \acp{pattern} extracted for the given \ac{re} type. The \ac{so} post mapping to the \ac{pattern} receiving the highest similarity score is then returned as the output. We now explain the different parts of our approach.

\subsection{Indexing of \ac{so} Posts}
\label{sec:clustering}

In this step we index \ac{so} posts by retaining only the posts that are related to fixing \acp{re}, and discard the rest. Our filtering criteria is that the post has: (1) at least one answer, (2) ``java" or ``android" tags, (3) \ac{re} type in the title, and (4) at least one parseable code snippet in the question. We define parseable code snippets as syntactically complete and conforming to Java grammar rules, as verified through any off-the-self parser, such as JavaParser~\cite{JavaParser}. The filtered \ac{so} posts are then grouped by \ac{re} type.

\subsection{\acf{abstract-tree}}
\label{sec:abstract-tree}

The \ac{abstract-tree} is a model capturing the structural and data relationships among the program statements of a given code snippet. Compared to traditional data structures, such as \ac{ast} and program dependence graph, the \ac{abstract-tree} focuses on simplifying, abstracting, and canonicalizing the low-level syntactic details. This is a more suitable data structure for our purpose, since it allows our approach to compare code despite the frequent differences in their syntax and semantics.

\subsubsection{\ac{abstract-tree} Definition}
\label{sec:abstract-tree-def}
Formally, we define the \ac{abstract-tree} as a directed graph of the form $\langle N, R_{s}, R_{d}, M\rangle$. Here $n \in N$ is a node in the graph that corresponds to a statement in the code. \Cref{fig:grammar} gives the grammar used for defining the nodes in the \ac{abstract-tree}. A node can be of type \code{method}, \code{loop}, \code{variable}, etc. Node type \code{root} designates the (unique) root nodes
of the \ac{abstract-tree}. Different node types further have different embodiments, such as \code{method} node is defined by three components, ``\code{caller}", which can be another node type (\eg variable or method), ``\code{name}", which is the method name, and a list of ``\code{arguments}", which can be a list of other nodes (\eg variable or literal). $R_{s} \subseteq N \times N$ is a set of directed edges, such that for each pair of nodes $\langle n_{1}, n_{2}\rangle \in R_{s}$, there exists a structural relationship between $n_{1}$ and $n_{2}$. Similarly, $R_{d}$ is a set of directed edges capturing the data relationship between $n_{1}$ and $n_{2}$. $M$ is a function $M: R_{d} \mapsto 2^{C}$ that maps each edge in $R_{d}$ to a set of tuples of the form $C: \langle v, \varphi\rangle$, where $v$ is a variable used in $n_{1}$ and $n_{2}$ and $\varphi$ is the type of $v$. (Details of building the \ac{abstract-tree} are discussed in \Cref{sec:abstract-tree-building}.) We define the following utility methods for a node $n$ in the \ac{abstract-tree}.

\begin{definition}
\label{def:construct}
\textit{construct ($n$): } returns the node type of $n$, such as \code{loop}, \code{if}, and \code{variable}. The only exception is for node types \code{declare}, \code{method} and \code{operation}, for which node type appended with the variable type, method name, and operator, respectively, are returned to make the constructs more specific.
\end{definition}

\begin{definition}
\label{def:types}
\textit{types ($n$): } returns the list of types for node $n$. For example, the node ``\code{method (Integer, reverse, \{x\})}" returns the list \{Integer, Integer\}, where \code{x} is an int variable.
\end{definition}

\setlength{\grammarparsep}{1px} 
\setlength{\grammarindent}{6.8em} 

\AtBeginEnvironment{grammar}{\small}
\begin{figure}
\begin{grammar}
<node> $\rightarrow$ <root> | <variable> | <method> | <if> | <loop> | <cast> | <arrAccess> | <fieldAccess> | <constructor> | <declare> | <literal> | <operation> | <instanceof> | <sync>

<method> $\rightarrow$ \texttt{method (}<caller>, <name>, \texttt{\{}<arguments>\texttt{\}})

<if> $\rightarrow$ \texttt{if (}<expression>\texttt{)}

<loop> $\rightarrow$ \texttt{loop (}<expression>\texttt{)}

<arrAccess> $\rightarrow$ \texttt{arrayAccess (}<caller> , <index>\texttt{)}

<cast> $\rightarrow$ \texttt{cast (}<type>, <expression>\texttt{)}

<constructor> $\rightarrow$ \texttt{constructor (}<type>, \texttt{\{}<arguments>\texttt{\})}

<declare> $\rightarrow$ \texttt{declare (}<variable>\texttt{)}

<fieldAccess> $\rightarrow$ \texttt{fieldAccess (}<caller>, <name>\texttt{)}

<literal> $\rightarrow$ \texttt{literal (}<type>, \texttt{value)}

<operation> $\rightarrow$ \texttt{operation (}<operator>, \texttt{\{}<operands>\texttt{\})}

<instanceof> $\rightarrow$ \texttt{instanceof (}<caller>, <type>\texttt{)}

<sync> $\rightarrow$ \texttt{synchronized (}<caller>\texttt{)}

<variable> $\rightarrow$ $\langle$ <name>, <type> $\rangle$

<caller> $\rightarrow$ <variable> | <method> | <fieldAccess> | <arrAccess> | `this' | `super'

<expression> $\rightarrow$ <variable> | <method> | <fieldAccess> | <arrAccess> | <operation> | <instanceof>

<arguments> $\rightarrow$ \texttt{(}<expression>\texttt{)*}

<operands> $\rightarrow$ <operand> \texttt{(','} <operand>\texttt{)*}

<operand> $\rightarrow$ <literal> | <expression>

<root> $\rightarrow$ `root'

<name> $\rightarrow$ variable name | method name | field name

<type> $\rightarrow$ primitive | class name | MISSING

<operator> $\rightarrow$ assign | binary | unary
\end{grammar}
\caption{\label{fig:grammar} Grammar of nodes in the \ac{abstract-tree}}
\end{figure}

\Cref{fig:buggy-code-abstract-tree} shows an excerpt of the \ac{abstract-tree} for the buggy code snippet in \Cref{fig:buggy-code}. The solid black arrows show the structural relationships and the dashed red arrows show the data relationships. 
Line numbers corresponding to the snippet are shown above the nodes.

\subsubsection{Building the \ac{abstract-tree}}
\label{sec:abstract-tree-building}
The process of building the \ac{abstract-tree} can be broken down into three steps: \textit{creation} of nodes and edges, \textit{attribution} of missing types, and \textit{abstraction} of constructs and types. 

\textit{Creation: } To build the \ac{abstract-tree} for a given code snippet, the approach first parses the snippet into an \ac{ast}. Nodes of the \ac{abstract-tree} are then created by traversing over the \ac{ast} and extracting groups of \ac{ast} nodes that can be simplified and summarized into one node of the \ac{abstract-tree}. For example, the sub-tree corresponding to the method call expression is summarized into the \ac{abstract-tree} node type ``method" with information about the method's caller, name, and arguments. The \ac{abstract-tree} retains the structural edges between the groups of summarized nodes obtained from the \ac{ast}. Data edges are then added to the \ac{abstract-tree} to capture the consecutive usage context of the variables used in the \ac{abstract-tree}. For example, the variable ``\code{fieldSchema}" is used in \code{getName()} node followed by its use in \code{remove()}. Hence, a data edge is added between the two nodes as shown in \Cref{fig:buggy-code-abstract-tree}.

\textit{Attribution: } After the \ac{abstract-tree} has been created, the attribution step augments the \ac{abstract-tree} by adding missing information about types. This step is necessary as the code snippets under consideration may not be complete, especially the ones coming from \ac{so}. Our approach is comprised of three kinds of attributions: (1) Use constructs to decide missing types of variables. For example, if the construct being used is \code{for-each}, then the type of the variable used in its iteration can be assigned as ``Collection", since the \code{for-each} loop typically operates over a collection object. Similarly, the type of a variable used to specify the conditions in the \code{if}, \code{while}, \code{do-while}, and \code{for} constructs can be assigned to be ``Boolean". (2) The caller of method invocations without explicit specifications can be presumed to be local methods, with the caller attributed as ``this". (3) Variables without explicit declaration of types, but containing assignments from literals, constructors, or arrays, are attributed with the types of the kinds of assignment. For example, type of \code{x} in the code snippet \code{x = 1;} is inferred to be ``Integer".

\textit{Abstraction: } This step abstracts and canonicalizes the information in the \ac{abstract-tree} to allow code comparison in the later steps to focus on high-level structure of the code rather than small syntactic differences. With this goal in mind, we designed the abstraction step to operate at different levels. First, canonicalize semantically equivalent constructs. We identify \code{for}, \code{for-each}, \code{while}, and \code{do-while} as semantic equivalents of each other, since they perform the same function despite the syntactic differences. For these  looping constructs, we canonicalize their node type as ``loop". For example, \code{for-each} at L216 in \Cref{fig:buggy-code} is translated to \code{loop} node in \Cref{fig:buggy-code-abstract-tree}. Similarly, \code{if}, \code{ternary operator (?:)}, and \code{switch-case} constructs are assigned a canonicalized representation of ``if" node. All binary operations (\eg \code{+} and \code{-}), unary operations (\eg \code{++}), and assignment of values to variables are generalized to an ``operation" node with the operator field specifying the operation type, such as PLUS, MINUS, and ASSIGN. Second, the abstraction step processes the data types of variables. Our approach first converts all primitive types into their corresponding wrapper classes, such as \code{int} is normalized to \code{Integer}. Then, the approach identifies user-defined or non-Java framework classes and re-attributes them as ``appClass" (type of fieldSchema is changed from FieldSchema to appClass in \Cref{fig:buggy-code-abstract-tree}). Lastly, the approach abstracts all collection class/interface types, such as \code{List} and \code{HashMap} into a common type ``Collection" (type of list object in \Cref{fig:so-code-abstract-tree} is changed from ArrayList to Collection). Normalization of the collection classes is driven by the observation that the root cause for the \ac{re} across all  different Collection classes is the same. Canonicalizing all collection classes allows our approach to focus on the exception causing scenario and find a relevant \ac{so} post addressing the root cause, rather than being tied up in type differences. Finally, the abstraction step scans for duplicate sub-trees within the \ac{abstract-tree}, assigns them to new variables, and refactors their use to point to the new variables. This refactoring helps reduce the overall size of the \ac{abstract-tree}. For example, the method chain, \code{table.getSd().getCols()}, appears at lines 216 and 218, hence, it is refactored into a new variable, \code{\$t1}, as shown in \Cref{fig:buggy-code-abstract-tree}.

\vspace{-0.25em}
\subsection{\ac{abstract-tree} Structural Alignment}
\label{sec:alignment}

We now describe our alignment component that is used in the \ac{localization} and code similarity matching steps. The goal of this component is to find the structural correspondence between two \acp{abstract-tree}. The \acp{abstract-tree} are first converted into \textit{structural trees} by removing their data edges. We then apply the APTED~\cite{APTED2016, APTED2015} tree-edit distance algorithm to find the corresponding nodes. The output is a set of tuples of the form $\langle n_{1}, n_{2}, M\rangle$, where $n_{1}$ and $n_{2}$ is a pair of matched nodes and $M$ indicates the type of match; full or partial.

Given a cost model, the APTED algorithm produces as output an optimal (minimal-cost) set of edit operations required for transforming structural tree $T_{1}$ into tree $T_{2}$. The different possible operations are: \textit{match}, \textit{delete}, \textit{insert}, and \textit{update}. We define the cost model as follows. The \textit{match} operation is designed to perform a leveled equivalence check between two nodes, $n_{1} \in T_{1}$ and $n_{2} \in T_{2}$. If both, the constructs (\Cref{def:construct}) and types (\Cref{def:types}), of $n_{1}$ and $n_{2}$ are identical, then it is considered as a \textit{full match}, and has no cost, implying the most preferred operation. If only the constructs match, then it is considered to be a \textit{partial match}, and the cost is 0.5. If nothing matches, then a unit cost is returned. The \textit{delete} operation generally has a unit cost, unless the node to be deleted is same as the node at the failing line, then it returns an \textit{infinite} 
cost. The intuition behind preempting the deletion of nodes at the failing line is to prevent false-positive alignment at locations in the code that are not related to the exception scenario. \textit{insert} and \textit{update} operations simply return a unit cost.

\subsection{\acf{localization}}
\label{sec:localization}

Given a \ac{so} post, the goal of this step is to identify lines in the question code snippet that are relevant to the exception scenario. To perform the \ac{localization}, our insight is that answer code snippets often suggest patches for the \ac{re} being discussed in the question and in doing so more directly reference the failure producing lines. Thus, our approach is to use the answer  snippets to identify the relevant lines and discard other lines. This idea of using answer code snippets is broadly analogous to spectrum-based fault localization techniques, which use failing test cases to locate the faulty lines.

The \ac{localization} step takes as input a \ac{so} post. For each question and answer code pair, the approach begins by representing them as \textit{\ac{abstract-tree}}$_{Q}$ and \textit{\ac{abstract-tree}}$_{A}$, respectively. It then aligns \textit{\ac{abstract-tree}}$_{Q}$ and \textit{\ac{abstract-tree}}$_{A}$ using the algorithm discussed in \Cref{sec:alignment} to obtain a set of matched nodes. Each matched node $n_{Q}$ $\in$ \textit{\ac{abstract-tree}}$_{Q}$, is annotated as relevant. After all of the answer code snippets in the post have been processed, \textit{\ac{abstract-tree}}$_{Q}$ is pruned by deleting any nodes not annotated through any answer.
This pruned \textit{\ac{abstract-tree}}$_{Q}$, called the \textit{\acf{pattern}}, is then produced as the output of the \ac{localization} step.

\subsection{Code Similarity Matching}
\label{sec:matching}

The goal of this step is find the degree of similarity between the developer's buggy code snippet and a \ac{so} post's question code snippet to determine the relevancy of the post for the developer. To determine relevance, our insight is to compare the two code snippets based on the similarity between their exception scenarios.

The code similarity matching step takes three inputs: the developer's buggy code, exception information (failing line and exception type), and \acp{pattern} obtained from the \ac{localization} of \ac{so} posts during the offline phase. The approach first converts the buggy code into \textit{\ac{abstract-tree}}$_B$. It then obtains the set, $P$, of \acp{pattern} corresponding to the given exception type. $p \in P$ is aligned with \textit{\ac{abstract-tree}}$_B$ using the algorithm described in \Cref{sec:alignment}. The approach then computes the similarity score (discussed below) for $p$ using the set of matched nodes given by the alignment function and analyzing \textit{\ac{abstract-tree}}$_B$ and $p$. The approach orders the \ac{so} posts corresponding to the \acp{pattern} in $P$ in descending order of similarity. Posts with same similarity score are ordered based on the number of user votes (high to low), which is an approximate indicator of the popularity of the post. Finally, the topmost \ac{so} post is returned as the output of the approach.

\vspace{-1em}
\begin{equation}
\mathit{similarity\_score} = w_{1} \times \mathcal{C} +
                            w_{2} \times \mathcal{T} +
							w_{3} \times \mathcal{V} + 
							w_{4} \times \mathcal{S}
\label{eq:simscore}
\end{equation}
\vspace{-1em}

\Cref{eq:simscore} shows the similarity score, which is a function of four weighted heuristics. The heuristics cover different aspects of similarity: structural (construct and type), data (var-use), and size (\ac{pattern}). Each of the four heuristics are normalized to report a value in the range [0, 1]. We now describe the heuristics.

\textbf{Construct similarity ($\mathcal{C})$} is the sum of \textit{partially} matched nodes divided by the total number of matched nodes. The partially matched nodes are obtained by scanning the set of tuples returned by the alignment algorithm where $M = \mathit{partial}$.

\textbf{Types similarity ($\mathcal{T}$)} is the sum of \textit{fully} matched nodes divided by the total number of matched nodes. Similar to above, fully matched nodes are obtained by scanning the tuples for $M = \mathit{full}$.

\textbf{Var-use similarity ($\mathcal{V}$)} measures the similarity between the uses of variables appearing in the matched nodes of $p$ and \textit{\ac{abstract-tree}}$_B$. The intuition behind this heuristic is that related exception scenarios should respect the same data relationships between their nodes as well. For a matched node pair, $\langle n_{p}, n_{B}\rangle$, the approach extracts the variables nodes at $n_{p}$ and $n_{B}$, respectively. For each variable, $v$, it then collects the set of nodes ($U_{v}$) where the variable is being used by traversing the data edges in the \ac{abstract-tree} under consideration. The approach then computes the Jaccard similarity between the $U_{v}$ of $p$ and that of \textit{\ac{abstract-tree}}$_B$. For example, for the matched  \code{loop} (...) nodes from \Cref{fig:buggy-code-abstract-tree} and \Cref{fig:so-code-abstract-tree}, $U_{\mathit{\$t1}}$ =  $U_{\mathit{list}}$ = \{\code{loop}, \code{remove}\}. Thus, their Jaccard similarity score is 1.0. Var-use similarity score is calculated as the sum of all Jaccard similarities of all variables in the matched nodes divided by the total number of variables. 

\textbf{\ac{pattern} size similarity ($\mathcal{S}$)} is given by \textit{min ($|p|$, $|p_{\mathit{ideal}}|$)} divided by \textit{max ($|p|$, $|p_{\mathit{ideal}}|$)}, where $|p|$ is the size (total number of nodes) of $p$ and $|p_{\mathit{ideal}}|$ is the size of the ideal \ac{pattern} for that exception type. This allows our approach to penalize rather small or large \acp{pattern}, which may contain too little or superfluous information that may lead to false positive matches. A small \ac{pattern} can be caused if the answer code snippets used in the \ac{localization} step do not fully capture the exception scenario, while a large \ac{pattern} can be caused by answer snippets that contain redundant matches with the question code snippet. Since knowing the ideal \ac{pattern} size for each \ac{re} context is not feasible, for a given exception type, we use \textit{median} size derived from all \acp{pattern} of the exception type as a proxy for the ideal size.

\section{Evaluation}
\label{sec:evaluation}

To assess the effectiveness of our approach, we conducted an empirical evaluation to address the following research questions:

\RQ{RQ1}{How effective is \toolname in recommending relevant \ac{so} posts for fixing \acp{re}?}

\RQ{RQ2}{How effective are the key contributions, \textit{localization}, \textit{matching}, and \textit{program abstraction}, in finding a relevant post?}

\RQ{RQ3}{How relevant are the \ac{so} posts suggested by \toolname compared to other state-of-the-art  techniques?}

\subsection{Implementation}
We implemented our technique as a Java prototype tool named \toolname. We used JavaParser (v3.14.3) \cite{JavaParser} to parse code snippets and build their \acp{ast}. We also minimally pre-processed the code snippets by adding enclosing class and/or method to improve their parseability. We leveraged the implementation of APTED algorithm~\cite{APTEDImpl, APTED2016, APTED2015} to compute the structural alignment between \acp{abstract-tree}. The four weights,  $w_{1\textrm{--}4}$, used in \Cref{eq:simscore} were set to 1, 2, 1, and 1.4, respectively. These weight values were selected by performing an empirical analysis on a subset of the dataset.

\subsection{Datasets}
\label{sec:datasets}
We instantiated \toolname on \ac{so} posts from the data dump released in March 2019~\cite{SODump}. Filtering by the criteria discussed in \Cref{sec:clustering} gave us a pool of 20,165 usable \ac{so} posts. The number of posts per exception type ranged from 3 to 10,920, with an average of 1,050 posts and a median of 109 posts.

For our evaluation, we created a benchmark of 78 instances extracted from the top 500 Java repositories on GitHub. Each instance is comprised of the buggy Java file throwing the \ac{re} and the failing line number. 
The top 500 GitHub repositories represent popular, large, and well-maintained projects. To select our evaluation instances, we first scanned the commit messages of the top 500 GitHub repositories for the mention of at least one of the 53 Java \ac{re} types in a fixing context. We established the fixing context by considering keywords such as \textit{fix}, \textit{resolve}, \textit{repair}, \textit{avoid}, and \textit{prevent}. We then filtered out duplicates, commits that did not contain any Java change files, and commits that were consisted of simply throwing the \ac{re}. 

After the filtration, we were left with 1,724 candidate patches across 19 unique \ac{re} types. We then categorized these candidates, by exception type, into four groups: \textit{high}, \textit{medium}, \textit{low}, and \textit{very low}, based on the frequency of their corresponding \ac{re} types. High category comprised of \ac{re} types having 100 or more candidates. Similarly, medium with 10--99 candidates, low with 2--9, and very low with only one candidate. This resulted in the four categories including 6, 5, 2, and 6 \ac{re} types, respectively. This distribution approximates the frequency of occurrence
of the different \acp{re} in the real-world. Finally, since our evaluation metrics involve manual inspection (\Cref{sec:metrics}), we chose to select a small but representative sample from each of the \ac{re} types. Thus, to mimic the geometric progression observed across the four \ac{re} categories, our methodology was to randomly select 8, 4, 2, and 1 candidates from the four categories, respectively, for each of the corresponding \ac{re} types. To collect the instance from each selected candidate, we downloaded the buggy Java file (\ie one version before the candidate's commit ID) and determined the failing line by analyzing bug reports, commit messages, and/or based on our domain knowledge. \Cref{tab:maestro} shows a summary of our evaluation dataset under the columns ``Category", ``\ac{re} type", and ``\# inst". The complete benchmark is available as part of the \toolname artifact
at~\cite{maestro-repo}.

\subsection{Evaluation Metrics}
\label{sec:metrics}
The goal of our evaluation is to measure the relevancy of the \ac{so} posts recommended by \toolname (RQ1), its variants (RQ2), and its competitors (RQ3) for fixing \acp{re}. However, there exists no ``ground truth" for calculating such a relevancy. This relevancy has to be derived through a manual inspection that is necessarily subjective. 

For this, we recruited two external participants to manually evaluate and provide the relevancy ratings for RQ1--3. Our participants were software professionals with a Java experience of 5--10 years. 
To further reduce bias, for each of the 78 instances, the participants were provided with eight \ac{so} posts produced by the different tools (1 by \toolname, 3 by its variants, and 4 by its competitors) in a \textit{randomized} and \textit{anonymized} fashion. The participants independently analyzed all of the instances, and then sat down together to resolve differences with one of the authors serving as a mediator to reach consensus~\cite{AROMA:OOPSLA2019, Monperrus-ESE2017}. Before resolving the differences, we measured inter-rater reliability using Cohen's Kappa~\cite{cohensKappa1960}, which gives a score of the degree of agreement between two raters. Across the 624 ratings provided by each participant, the value of Kappa ($\kappa$) was found to be 0.63, implying substantial agreement between the raters~\cite{cohensKappaInterpretation1977}.

The participants were asked to rate each \ac{so} post in one of the following four categories: \textbf{Instrumental (I)}: The participant feels confident that the \ac{so} post captures the \ac{re} scenario precisely, and offers a highly effective repair in the context of the instance.
\textbf{Helpful (H)}: The participant finds the \ac{so} post informative, \ie offers insight into the \ac{re}, but does not provide a direct solution for fixing the \ac{re} in the context of the instance.
\textbf{Unavailable (U)}: No \ac{so} post was recommended by the tool. In a real-world deployment this would have required the end-user to perform their own manual search for a solution. 
\textbf{Misleading (M)}: The participant finds the \ac{so} post highly irrelevant to the instance.

This rating scale broadly follows the approach of Zimmermann et al.~\cite{scaleSelection2015, scaleSelection2014}. We did not include a ``Don't Know" category following the advice of Kitchenham et al.~\cite{metricsGuidelines2008}, as our participants were well-experienced to always make an informed decision.

To characterize the overall effectiveness in recommending a relevant \ac{so} post, we used the following metrics, again following prior research~\cite{scaleSelection2015, scaleSelection2014}, where they have been shown to be successful in avoiding scale violations~\cite{metricsGuidelines2008}. 

\begin{table}[!htbp]
  \centering
  \label{tab:metrics}\setcellgapes{3pt}\makegapedcells
  \begin{tabular}{p{5cm}|p{2.8cm}}
    \toprule
    \textbf{I-score}: Percentage of \textit{perfect} \ac{so} posts, \ie rated Instrumental & $\textit{I-score} = \frac{I}{I + H + U + M}$\\
    \midrule
    \textbf{IH-score}: Percentage of \textit{relevant} \ac{so} posts, \ie rated Instrumental or Helpful & $\textit{IH-score} = \frac{I + H}{I + H + U + M}$\\
    \midrule
    \textbf{M-score}: Percentage of \textit{irrelevant} \ac{so} posts, \ie rated Misleading & $\textit{M-score} = \frac{M}{I + H + U + M}$\\
    \bottomrule
  \end{tabular}
\end{table}

Details of the participant ratings for the 78 instances across all eight tools (RQ1--3) are available at~\cite{maestro-repo}

\begin{table}[t]\scriptsize
\centering
\caption{Effectiveness Results of \toolname}
\label{tab:maestro}
\begin{tabular}{cl@{\hskip 3pt}r|rrr}
\toprule
{\bf Category} & {\bf \ac{re} type} & {\bf \# inst} & {\bf I-score} & {\bf IH-score} & {\bf M-score}\\
\midrule
\multirow{6}{*}{High} & ClassCastException & 8 &  0.63 & 1.00 & 0.00\\
\multirow{6}{*}{[100, $\infty$]}& ConcurrentModificationException & 8 & 0.75 & 1.00 & 0.00\\
& IllegalArgumentException & 8 & 0.38 & 0.50 & 0.38\\
& IllegalStateException & 8 & 0.25 & 0.50 & 0.25 \\
& IndexOutOfBoundsException & 8 & 0.38 & 0.63 & 0.38\\
& NullPointerException & 8 & 0.00 & 0.38 & 0.63\\
\midrule
\multirow{5}{*}{Medium} & ArithmeticException & 4 & 1.00 & 1.00 & 0.00\\
\multirow{5}{*}{[10, 99]} & NoSuchElementException & 4 & 0.25 & 0.25 & 0.75\\
& RejectedExecutionException & 4 & 0.75 & 1.00 & 0.00\\
& SecurityException & 4 & 0.25 & 0.75 & 0.25\\
& UnsupportedOperationException & 4 & 0.00 & 0.75 & 0.25\\
\midrule
Low & EmptyStackException & 2 & 0.50 & 1.00 & 0.00\\
{[}2, 9{]} & NegativeArraySizeException & 2 & 0.50 & 0.50 & 0.50\\
\midrule
\multirow{6}{*}{Very Low} & ArrayStoreException & 1 & 0.00 & 1.00 & 0.00\\
\multirow{6}{*}{[1, 1]} & BufferOverflowException & 1 & 0.00 & 1.00 & 0.00\\
& BufferUnderflowException & 1 & 0.00 & 0.00 & 1.00\\
& CMMException & 1 & 0.00 & 1.00 & 0.00\\
& IllegalMonitorStateException & 1 & 0.00 & 1.00 & 0.00\\
& MissingResourceException & 1 & 1.00 & 1.00 & 0.00\\
\midrule
\multicolumn{3}{r}{\textbf{Overall}} & \textbf{0.40} & \textbf{0.71} & \textbf{0.26}\\
\bottomrule
\end{tabular}
\end{table}

\subsection{RQ1: Effectiveness of \toolname}
\label{sec:rq1}
\Cref{tab:maestro} shows the results grouped by \ac{re} types. \toolname reported an overall IH-score of 71\%, \ie \ac{so} posts for 55 out of 78 instances were rated \textit{relevant} (Instrumental or Helpful). Of these, 31 posts were rated Instrumental (I-score = 40\%). This shows that \toolname was effective in recommending relevant \ac{so} posts for fixing \acp{re}.

A perfect IH-score (100\%) was reported for 10 out of 19 \ac{re} types. One prominent pattern that we observed among these successful cases was that the exception scenario was typically well-defined with a specific sequence of actions leading to the \ac{re}. An example of such a pattern is described in \Cref{sec:motivating-example} for \code{ConcurrentModification}. It is a multi-line pattern with a distinctive \textit{structural} dependency --- \code{remove()} enclosed within a \code{loop} --- and \textit{data} dependency --- same \code{collection} object is used in \code{loop} iteration and \code{remove()} invocation. Another example of a well-defined, but single line pattern is shown in \Cref{fig:inst-rq1-eg2}. \code{NoSuchElementException} is thrown when the \code{Iterator} has no more elements. \toolname searches through a pool of over 700 posts for this \ac{re} type to find the post shown in \Cref{fig:so-rq1-eg2}. The post is highly relevant as it poses a similar problem as the buggy code. However, finding such an accurate post manually can prove to be rather challenging, as was reported by our user study participants (\Cref{sec:user-study}), who particularly appreciated \toolname's post by saying it was \textit{``better than what they found"} and that \textit{``the post can solve the problem perfectly"}. The abstraction encoded in \toolname's \ac{abstract-tree} 
coupled with its matching algorithm facilitates anchoring on this post: the expression, \code{iterator().next()}, matches \textit{fully} while \code{visibleTiles.keySet()} is found to match \textit{partially} after abstracting out the details.

\begin{figure}[t]
\centering
\begin{subfigure}[b]{\columnwidth}
\begin{lstlisting}[language=Java, framexleftmargin=0.8em, xleftmargin=1.1em, numbersep=3pt, basicstyle=\fontsize{7pt}{7pt}\selectfont\ttfamily, numberstyle=\scriptsize]
...
if (name == null) {
  return @\colorbox{blue}{\strut uploaders.values()}@@\colorbox{pastelyellow}{\strut .iterator().next();}@ @\hspace{0pt} {\normalsize \textbf{\color{red} $\leftarrow$ RE}}@
}
...
\end{lstlisting}
\vspace{-5pt}
\caption{\label{fig:inst-rq1-eg2}Buggy Code from bazelbuild/bazel}
\end{subfigure}
\par\bigskip
\begin{subfigure}[b]{\columnwidth}
\begin{lstlisting}[language=Java, framexleftmargin=0.8em, xleftmargin=1.4em, numbersep=3pt, basicstyle=\fontsize{7pt}{7pt}\selectfont\ttfamily, numberstyle=\scriptsize]
...
tile = @\colorbox{blue}{\strut visibleTiles.keySet()}@@\colorbox{pastelyellow}{\strut .iterator().next();}@
if (tile != null) {
    ...
}
...
\end{lstlisting}
\vspace{-5pt}
\caption{\label{fig:so-rq1-eg2}\ac{so} post: \#13053195}
\end{subfigure}
\caption{\label{fig:rq1-eg2}Relevant post for NoSuchElementException\\
(\colorbox{pastelyellow}{\strut Yellow} shows \textit{full match} and \colorbox{blue}{\strut Blue} shows \textit{partial match})}
\end{figure}

\begin{figure}[t]
\centering
\begin{subfigure}[b]{0.54\columnwidth}
\begin{lstlisting}[language=Java, framexleftmargin=0.8em, xleftmargin=1.1em, numbersep=3pt, basicstyle=\fontsize{7pt}{7pt}\selectfont\ttfamily, numberstyle=\scriptsize]
...
@\colorbox{blue}{\strut while }@(!isShutdown.get()) @\hspace{0pt} {\normalsize \textbf{\color{red} $\leftarrow$ RE}}@
 ...
  if (@\colorbox{blue}{\strut this.metricUploader}@ != null)
...
\end{lstlisting}
\vspace{-5pt}
\caption{\label{fig:inst-rq1}Buggy Code from alibaba/jstorm}
\end{subfigure}
\begin{subfigure}[b]{0.44\columnwidth}
\begin{lstlisting}[language=Java, framexleftmargin=0.8em, xleftmargin=1.4em, numbersep=3pt, basicstyle=\fontsize{7pt}{7pt}\selectfont\ttfamily, numberstyle=\scriptsize]
...
@\colorbox{blue}{\strut for}@ (Square[] s:gameBoard)
  for (Square ss : s)
     ss = @\colorbox{blue}{\strut Square.EMPTY}@;
...
\end{lstlisting}
\vspace{-5pt}
\caption{\label{fig:so-rq1}\ac{so} post: \#21819264}
\end{subfigure}
\caption{\label{fig:rq1-eg}Irrelevant post for NullPointerException \\
(\colorbox{blue}{\strut Blue} shows \textit{partial match})}
\end{figure}

On the contrary, \ac{re} types such as \code{NullPointerException} showcase an overly generic pattern, viz. dereferencing a null object, which can have a wide range of manifestations. This makes it challenging for \toolname to anchor upon specific program elements that can help find the best post. \Cref{fig:rq1-eg} shows an example of this case. The \ac{re} is thrown at line 2 in the buggy code because ``\code{get()}" is invoked on ``\code{isShutdown}", which is null. Since the program elements are not specifically pertinent to the \ac{re}, \toolname finds an arbitrary post that matches at irrelevant constructs, such as \code{loop} (line 2) and \code{field access} (line 4). Our investigations revealed two other reasons when \toolname reported irrelevant posts. First, inaccuracies in the \ac{localization} step lead to under or over specific \acp{pattern}. This happens when the answers  do not capture the failure scenario succinctly \textit{and} with a sufficient code context. As a consequence of this \toolname 
again anchors on irrelevant program elements, resulting in arbitrary posts. The second reason is when the exception scenario in the buggy code is very rare or application specific, \toolname may not report any post, which happened for 3 out of the 78 instances, or it may report irrelevant posts based on peripheral matches.

In our experiments, we found that \toolname takes a median 2.6 sec (average = 76 sec) end-to-end to find the most relevant \ac{so} post on a 8-core desktop machine. We believe this makes \toolname effective for real-time use.

\subsection{RQ2: Key contributions of \toolname}
\label{sec:rq2}
\subsubsection{Design} In this experiment, we evaluate the importance of the three main contributions of our work by creating three baseline variants of \toolname. The first variant, \textsc{\toolname-NoLoc}, measures the impact of the \ac{localization} step (\Cref{sec:localization}) by removing it from the workflow. The second variant, \textsc{\toolname-SimpleMatch}, assesses the importance of the heuristics-based weighted similarity score computation (\Cref{sec:matching}) by replacing it with a simple sum of the matched nodes (full and partial), \ie the post with the highest number of matches is returned as the most relevant post. Lastly, the third variant, \textsc{\toolname-AST}, evaluates the importance of the \ac{abstract-tree} representation (\Cref{sec:abstract-tree}) by replacing it with \ac{ast}. For this variant, we implemented the cost function for structural alignment (\Cref{sec:alignment}) as follows: A strict syntactic match between method calls, class names, data types, etc. constitutes a \textit{full} match, while matches between rule nodes, such as \code{IfStatement} and \code{MethodCallExpr}, constitute a \textit{partial} match.

\subsubsection{Results}
\begin{figure}[t]
  \centering
    \includegraphics[width=\columnwidth]{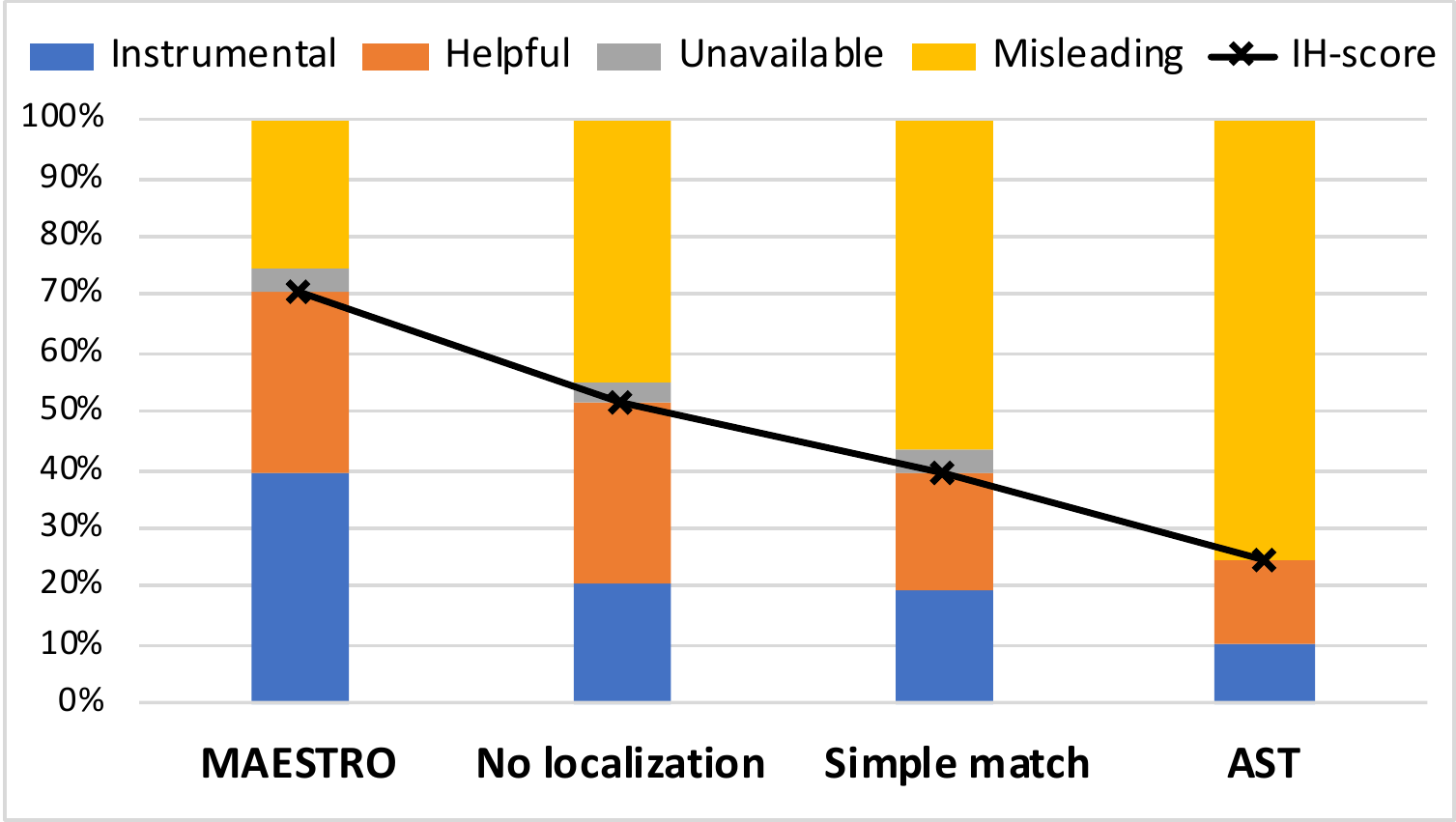}
    \caption{\toolname vs. its variants}
    \label{fig:variants-chart}
\end{figure}

\Cref{fig:variants-chart} shows the distribution of the participant ratings for \toolname and its variants. The IH-scores for the three variants show a significant almost-linear drop compared to \toolname: 28\%, 44\%, and 66\%, respectively. The primary reason for this decrease is that the variants tend to match with the lines in the \ac{so} snippet that are not related to the exception scenario. Concretely, \textsc{\toolname-NoLoc} matches entire \ac{so} snippets with the instance. This falsely causes the snippets to match peripherally to the core exception scenario, leading them to be ranked higher. Conversely, while \textsc{\toolname-SimpleMatch} employs \ac{localization}, it gives preference to \acp{pattern} that match maximally with the instance, albeit incorrectly. \toolname overcomes this problem by penalizing inflated \acp{pattern} by comparing them with the ideal \ac{pattern} size, as discussed in \Cref{sec:matching}. On the other hand, while \textsc{\toolname-AST} employs both \ac{localization} and the heuristics-based similarity scoring, it performs poorly because the \ac{ast} representation alignment results in very few \textit{full} matches, while a high number of false positive \textit{partial} matches. Overall, this experiment demonstrates the strengths of the different components in \toolname, 
and how they contribute to its ability to meaningfully compare bug scenarios and thereby identify
relevant \ac{so} posts.

\begin{figure}[t]
  \centering
    \includegraphics[width=\columnwidth]{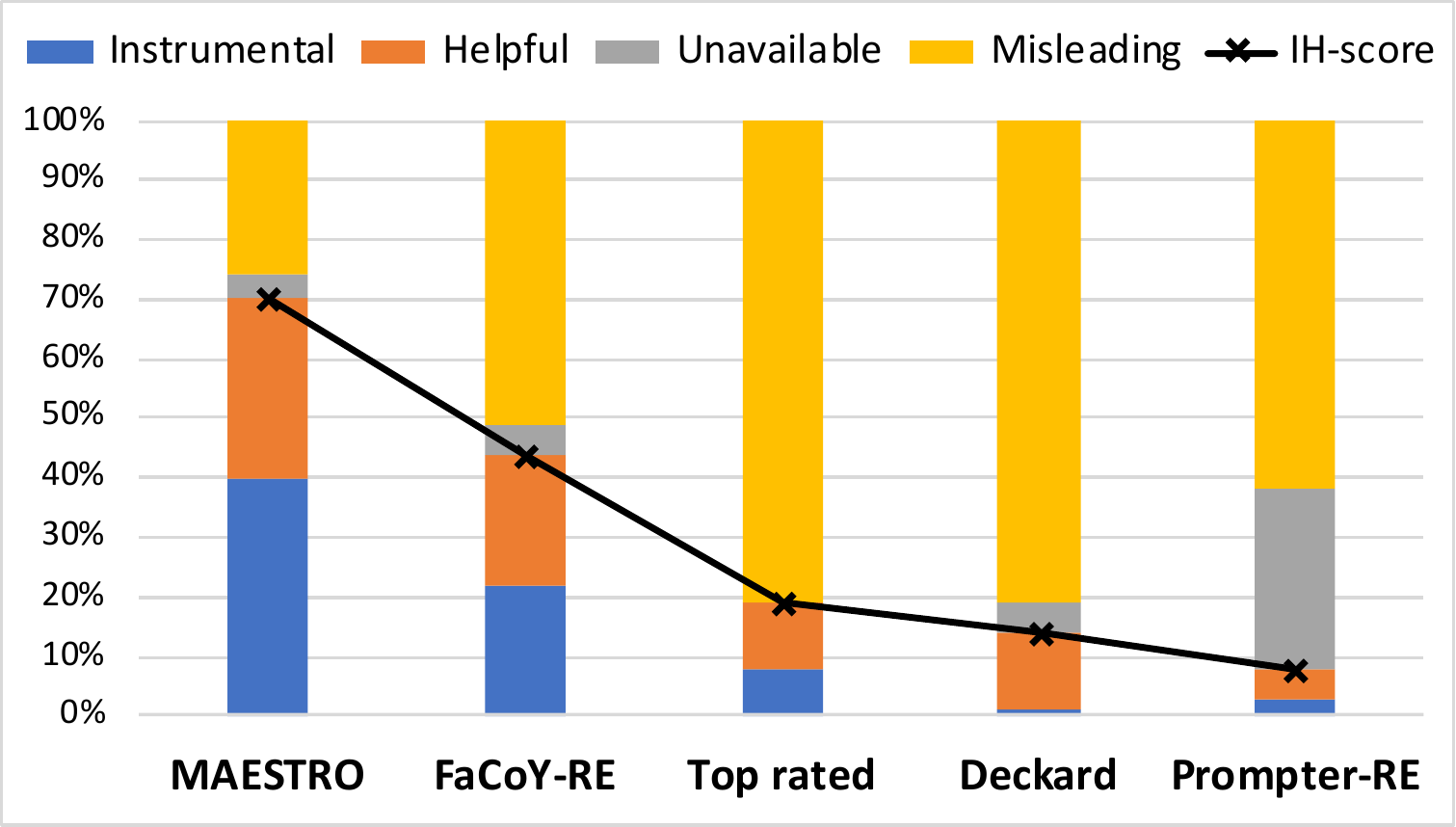}
    \caption{\toolname vs. other techniques}
    \label{fig:competitors-chart}
\end{figure}

\subsection{RQ3: Comparison with other techniques}
\subsubsection{Design} \toolname does not have any direct competitors, since there exist no techniques targeting its exact use case. However, state-of-the-art (1) \ac{so} recommendation techniques, (2) code matching techniques, and (3) the native \ac{so} search engine, can serve as viable alternatives. Representing the first group is Prompter~\cite{Prompter:MSR2014}, the only existing technique that searches and suggests \ac{so} posts for a given context, albeit from a code completion perspective. To adapt to our use case, we created \textbf{Prompter-RE}, which prefixes Prompter's default search query with the \ac{re} name~\cite{searchOrder}. To provide sufficient code context, Prompter-RE is given the entire method from the instance containing the failing line, as its input. For the second group, we used state-of-the-art syntactic and semantic code clone detectors to match input buggy code with \ac{so} code snippets. For the syntactic code clone detector, we selected \textbf{Deckard}~\cite{DECKARD:ICSE2007} which is based on comparing \acp{ast}. We made this choice by comparing its performance with the current-generation token-based detector, SourcererCC~\cite{SourcererCC:ICSE2016}. Details of this experiment can be found at \cite{maestro-repo}. Aroma's~\cite{AROMA:OOPSLA2019} light-weight search could also have served as an alternative syntactic code similarity detector. However, since principally Deckard and Aroma both capture and compare syntactic and structural relationships in code, we selected Deckard as a representative. As our semantic code clone detector, we used FaCoY~\cite{FaCoY:ICSE2018}. We compare \toolname with the first phase of FaCoY, which produces a ranked list of \ac{so} posts by matching structural code tokens. To suit our use case, we created \textbf{FaCoY-RE} by changing its default implementation to compare input with \ac{so} \textit{question} snippets, rather than answers. We instantiated both Deckard and FaCoY-RE on the same \ac{so} pool as \toolname, and provided as input a fragment of the instance encompassing the failing line and its immediate surrounding context, $\pm$3 lines~\cite{elixir2017}. Lastly, we compare with the native \ac{so} search engine~\cite{so-search-engine}, which for a given \ac{re} type keyword, returns a list of \textbf{top rated} posts, ranked by relevance.

\subsubsection{Results}
\Cref{fig:competitors-chart} shows the distribution of participant ratings for the 78 instances across different competitors. The IH-scores of FaCoY-RE, Top rated, Deckard, and Prompter-RE are 44\%, 19\%, 14\%, and 8\%, respectively. \toolname outperforms all the techniques with a substantial margin, showing 38\%--89\% improvement in IH-score.

FaCoY-RE finds \ac{so} posts using tf-idf and Cosine similarity, which calculates the highest overlap between less frequent code tokens. This does not work in cases where the code elements are commonly occurring ones, such as in \code{IndexOutOfBoundsException}, or in cases with extremely rare (\eg app specific) tokens that may not have any representation on \ac{so}. Deckard demonstrated a similar problem in such instances, where finding syntactic matches between buggy code and \ac{so} snippets resulted in irrelevant or no matches. This underscores the strength of \toolname's \ac{abstract-tree} representation which is able to sufficiently abstract out the syntactic details to compare at a high-level. The main reason behind low quality posts reported by Prompter-RE is that its technique is based on a bag of words algorithm that tries to find a general-purpose post discussing the overall function of the buggy code, rather than focusing on the exception scenario. We also observed that although we had augmented the search query, the added \ac{re} name was in some cases removed by the search engine (\eg Google), likely due to the close relevancy among the other words in the query. In such situations, the posts returned by Prompter-RE did not even pertain to the \ac{re}. Lastly, although the \ac{so} top-rated post for a given \ac{re} type contains rich and informative content, it is unable to provide satisfactory resolution for \emph{all} different exception scenarios of the \ac{re}.

This experiment shows that state-of-the-art techniques, applied in a straight-forward manner, may not be well-suited for \toolname's use case of finding most relevant \ac{so} posts for fixing \acp{re}.

\subsection{User Experience Case Study}
\label{sec:user-study}

We conducted a controlled study (results at~\cite{maestro-repo}) to understand developers' experience using \toolname. We presented 10 instances from our dataset to 10 Java developers, with each instance evaluated by five different participants. For a given instance, the participants were asked to manually search and report a relevant \ac{so} post. Then they were shown and asked to rate the post recommended by \toolname, and provide feedback about their experience. We decided not to measure time saving, as other research has shown that manual time varies greatly by experience level and has little correlation with task complexity~\cite{AROMA:OOPSLA2019}.

\toolname met with a high user approval with the IH-score of 80\% (I-score: 50\%). Comments left by the participants were very insightful in understanding their reasoning behind the qualitative assessment of the posts. We found that broadly the specific characteristics of the post itself, as well as, the participants' manual search influenced their judgement of \toolname's post. For the posts rated \textbf{instrumental}, the participants provided feedback, such as \textit{``Better post than I found from my search"}, \textit{``I found the exact one searching"}, \textit{``Post can solve the problem perfectly"}, and \textit{``The concept and solution of the post is correct"}. For posts rated \textbf{helpful}, example comments were \textit{``Get basic understanding of the potential reason for the error"}, \textit{``It doesn't solve the problem, but the discussion is relevant"}, and \textit{``Provides javadoc and reasonable understanding to peruse why problem was happening"}. While for posts marked \textbf{misleading}, comments were \textit{``Same exception, but totally different context"} and \textit{``The post is not really addressing the real problem"}. Overall, such quality attributions strongly resonate with our observations discussed in \Cref{sec:rq1}. We found that the participants generally had a positive attitude towards \toolname, as it was able to find posts of comparable or even higher quality than them, for most instances.

\subsection{Limitations}
\textbf{Mining \ac{so}:} Developer forums like \ac{so} only discuss commonly occurring, generic development issues. \toolname inherits this limitation in that it cannot assist with very application-specific \acp{re}, \eg an \ac{re} rooted in the semantics of application-specific APIs. It is also not very helpful with overly generic \acp{re} like \code{NullPointerException} where there is no common, yet sufficiently descriptive pattern to the exception, and a specific post discussing it.

\textbf{Code-based search:} \toolname's current search relies exclusively on an analysis of code snippets in the Q\&A threads. Although this provides relevant matches in a significant fraction of instances, mining and incorporating information from the natural language text in the posts would be a valuable next step.

\textbf{Scope of analysis:}  Our \ac{localization} step uses the structure of answer snippets to pin-point the failure-inducing statements. However, posts with lengthy and/or non-specific answer snippets may lead to sub-optimal localization and match results. In future work we propose to use other sources of information, including static or dynamic analysis of question and answer snippets or information in the surrounding text to improve the accuracy of localization. Another limitation is that currently \toolname's analysis can only handle intra-procedural failure scenarios. This design is driven by the observation that the vast majority of \ac{re} scenarios tend to be succinct and local. However, extend \toolname to support to inter-procedural scenarios could further expand its scope.

\textbf{Construct Validity}: Judging the relevance of \ac{so} posts produced by any of the tools (Section~\ref{sec:evaluation}: RQ1-RQ3) is an inherently subjective task, and hence a potential threat. We mitigated this threat by specifying clear criteria for each of a 4-valued rating scale, consistent with previous work~\cite{scaleSelection2015, scaleSelection2014, metricsGuidelines2008}. Further, we used two \emph{non-authors} to independently rank each output instance (post), to remove author-bias and get a plurality of opinions. Then we used a discussion process for the raters to reach consensus on instances of different opinion, also following previous work~\cite{Monperrus-ESE2017, AROMA:OOPSLA2019}. Finally, we calculate and report Cohen's Kappa~\cite{cohensKappa1960}, a measure of inter-rater reliability, showing substantial agreement between the raters' original ratings.

\vspace{-1em}

\section{Related Work}
\textbf{Mining Q \& A sites.}
The work most closely related to ours is \prompter~\cite{Prompter:MSR2014, Seahawk:CSMR2013} which continuously searches and recommends relevant \ac{so} posts to a developer as she develops code in an IDE. \prompter models the user's code as a bag of words, creates a search query of rare tokens derived from it, and uses this query to search \ac{so}. Libra~\cite{Libra:ICSE2017} augments this approach by including relevant terms from the user's recent browsing history. However, while these techniques are a good fit for the general ``code completion'' use-case that they target, as shown in Section~\ref{sec:evaluation}, they do not work well for our use-case where more complete code and specific (\ac{re}) error information is available. Hence  our approach performs a more structured comparison of the user's code with the \ac{so} code snippets, rather than using a bag-of-words model.

In recent work Zhang et al.\cite{ExampleStack:ICSE2019} propose a tool {\sc ExampleStack}, to guide developers in adapting code snippets from relevant \ac{so} posts to their own codebase. {\sc ExampleStack} nicely complements our contribution of finding the relevant posts. {\sc CSnippEx}~\cite{CSNIPPEX:ISSTA2016} proposes an approach to make \ac{so} code snippets compilable by adding missing imports and variable, method, and class declarations. Such a technique can make a larger fraction of \ac{so} code analyzable by approaches such as ours. Nagy et al.~\cite{Nagy:ICSME2015} mine common error patterns in SQL queries for potential use by SQL developers. SOFix~\cite{SOFix:SANER2018} mines \ac{so} to \emph{manually} extract a set of repair schemas for use in a generate-and-validate automatic program repair (APR) tool. In both these works the aim is \emph{offline} mining of \ac{so} for subsequent use in specialized use-cases. By contrast, our problem is real-time recommendation of relevant \ac{so} posts on an instance-specific basis. QACrashFix~\cite{QACrashFix:ASE2015} uses error information in an Android-related crash bug to gather a population of relevant \ac{so} posts and then uses the posts' answer code snippets in a generate-and-validate APR approach to fix the bug. Unlike us, here the core contribution is on the use of information in \ac{so} posts for repair, rather than the search for relevant posts. 

\textbf{Code clone detection and code search.} 
Syntactic code clone detection techniques detect syntactically similar code fragments (\ie Type 1,2,3 clones) by matching tokens, such as CCFinder~\cite{CCFinder:TSE2002} and {\sc SourcererCC}~\cite{SourcererCC:ICSE2016}, or comparing \acs{ast}s, such as {\sc Deckard}~\cite{DECKARD:ICSE2007}, or using hybrid approaches such as NiCad~\cite{NiCad:ICPC2011}. 
However, as discussed in Sections~\ref{sec:introduction},\ref{sec:motivating-example} and empirically evaluated in Section~\ref{sec:evaluation} matching the developer's code with the \ac{so} code snippet is not a typical syntactic code clone detection problem because of the degree of dissimilarity between the two. Semantic code clone detection (\ie Type 4 clones) match syntactically dissimilar but semantically equivalent code fragments. Prominent representatives include the work by White et al.~\cite{White:ASE2016} based on deep learning, Oreo~\cite{Oreo:FSE2018} which combines information retrieval, machine learning, and software metrics,  CCAligner~\cite{CCAligner:ICSE2018} which specializes in large-gapped clones, and FaCoY~\cite{FaCoY:ICSE2018} which uses a novel \emph{query alternation} strategy leveraging natural language descriptions of code. Our problem is not suitable for these techniques either, since our problem instances do in fact possess structural similarity, \emph{provided} a technique can localize the exception-triggering segments and suitably abstract away syntactic differences. Our technique is tailored to do precisely this. Code-to-code search engines such as Aroma~\cite{AROMA:OOPSLA2019} and Krugle~\cite{Krugler2013} form another related body of work. However, since their purpose is code recommendation for the purpose of code completion or code enhancement they seek to find extensions or modifications of the query code, rather than seeking to align the error-producing scenarios of the two code segments, as we do.

\textbf{Debugging, patching, and recovery for runtime errors.} Our work is inspired by the body of research on remediation of runtime errors. Sinha et al.~\cite{Sinha:ISSTA2009} proposed one of the earliest techniques for fault localization and repair of Java runtime exceptions, NPEFix~\cite{NPEFix:arXiv2015} proposes a generate-and-validate APR approach for null-pointer exceptions (NPE), VFix~\cite{VFix:ICSE2019} uses data and control-flow analysis to prune the repair space for NPEs and generate more accurate repairs, and Genesis~\cite{Genesis:FSE2017} automatically extracts repair patterns specific to various exception types to use in an APR approach. There is also interesting research on isolating and recovering from runtime errors~\cite{Long:PLDI2014, Ares:ASE2016}. 
The above techniques use program analysis as the basis for their remediation. By contrast, our work facilitates the use of crowd-sourced knowledge available in online Q\&A sites for this purpose, and is therefore complementary to the above.

\section{Conclusion}
In this work we presented a technique and prototype tool called \toolname to automatically recommend an \ac{so} post most relevant to a given Java \ac{re} in a developer's code. Specifically, \toolname returns the post best matching the exception-generating program scenario in the developer's code.
To extract and compare the exception scenario, \toolname first uses the answer code snippets in a post to implicate relevant lines in the post's question code snippet and then compares these lines with the developer's code in terms of their respective \acf{abstract-tree} representations. The \ac{abstract-tree} is a simplified and abstracted derivative of an \ac{ast} that enables an effective high-level semantic comparison, while discarding low-level syntactic or semantic differences. An evaluation of \toolname on a benchmark of \benchmarkSize instances of Java \acp{re} extracted from the top 500 Java GitHub projects showed that \toolname can return a relevant \ac{so} post for \maestroIHScore of the exception instances, compared to relevant posts returned in only \otherIHScoreLow \,- \otherIHScoreHigh instances, by four competitor tools based on state-of-the-art techniques.  Further, in a user experience study of \toolname with 10 Java developers, the participants judged \toolname as reporting a highly relevant or somewhat relevant post in \uStudyMaestroIHScore of the instances, and in some cases, even better than the one manually found by the participant.

\section{acknowledgments}
We would like to express our gratitude for the significant time and effort invested by our two participants, who evaluated several hundred \ac{so} posts and provided the relevancy ratings for RQ1--3. We also thank the participants of our user experience case study for their valuable feedback. Finally, thanks to the anonymous reviewers for their constructive input, which helped improve the quality of this manuscript.
 
\bibliographystyle{ACM-Reference-Format}
\bibliography{runtime}

\end{document}